%% file: main.tex
\pdfoutput=1

\PassOptionsToPackage{sort}{natbib}

\documentclass[11pt]{article}
\usepackage[final]{acl}

\input{preamble/packages}
\input{preamble/definitions}
\input{preamble/authors}

\title{Bootstrapped Pre-training with Dynamic Identifier Prediction \\ for Generative  Retrieval}

\begin{document}

\maketitle

\begin{abstract}
Generative retrieval uses differentiable search indexes to directly generate relevant document identifiers in response to a query. 
Recent studies have highlighted the potential of a strong generative retrieval model, trained with carefully crafted pre-training tasks, to enhance downstream retrieval tasks via fine-tuning.
However, the full power of pre-training for generative retrieval remains underexploited due to its reliance on pre-defined static document identifiers, which may not align with evolving model parameters. 
In this work, we introduce BootRet, a \underline{boot}strapped pre-training method for generative \underline{ret}rieval that dynamically adjusts document identifiers during pre-training to accommodate the continuing  memorization of the corpus. 
BootRet involves three key training phases: 
(i) initial identifier generation, 
(ii) pre-training via corpus indexing and relevance prediction tasks, and 
(iii) bootstrapping for identifier updates. 
To facilitate the pre-training phase, we further introduce noisy documents and pseudo-queries, generated by large language models, to resemble semantic connections in both indexing and retrieval tasks. 
Experimental results demonstrate that BootRet significantly outperforms existing pre-training generative retrieval baselines and performs well even in zero-shot settings.
\end{abstract}

\input{sections/introduction}
\input{sections/related_work}
\input{sections/method}

\input{sections/experimental_settings}
\input{sections/experimental_resutlts}

\input{sections/conclusion}

\clearpage

\input{sections/limitations}

\bibliography{references}

\appendix
\input{sections/appendix}

\end{document}

%% file: preamble/packages.tex
\usepackage[final]{acl}
\usepackage{times}
\usepackage{latexsym}
\usepackage[T1]{fontenc}
\usepackage[utf8]{inputenc}
\usepackage{microtype}
\usepackage{inconsolata}

\usepackage{multirow}
\usepackage{multicol}
\usepackage{bbm}
\usepackage{dsfont}
\usepackage[inline]{enumitem} 
\usepackage{amsmath}
\usepackage{mathtools}
\usepackage{url}            
\usepackage{csquotes}
\usepackage{natbib}
\usepackage{graphicx}
\usepackage{acronym}
\usepackage{newfloat}
\usepackage{listings}
\usepackage{algorithm}
\usepackage{algorithmic}
\usepackage{booktabs}
\usepackage{xcolor}

\usepackage[skip=5pt]{caption}

%% file: preamble/definitions.tex
\acrodef{GR}{generative retrieval}
\acrodef{MLE}{maximum likelihood estimation}
\newcommand{\heading}[1]{\vspace*{.5mm}\noindent\textbf{#1.}}

%% file: preamble/authors.tex
\author{Yubao Tang$^{1,2}$ \space \space \textbf{Ruqing Zhang}$^{1,2}$  \space \space  \textbf{Jiafeng Guo}$^{1,2}$\thanks{Corresponding author.} \space \space  \textbf{Maarten de Rijke}$^{3}$  \space \space \\ \textbf{Yixing Fan}$^{1,2}$ \space \space  \textbf{Xueqi Cheng}$^{1,2}$\\
        $^1$CAS Key Lab of Network Data Science and Technology, ICT, CAS \\ 
         $^2$University of Chinese Academy of Sciences\\
         $^3$University of Amsterdam\\
         \texttt{\{tangyubao21b,zhangruqing,guojiafeng,fanyixing,cxq\}@ict.ac.cn}\\
         \texttt{m.derijke@uva.nl}}

%% file: sections/introduction.tex
\section{Introduction}
Document retrieval is an important task with widespread applications, such as question answering \cite{karpukhin2020dense,DBLP:conf/acl/LeeCT19} and fact verification \cite{chakrabarty2018robust,olivares2023enhancing}, which aims to retrieve candidate documents from a huge document collection for a given query \cite{gao2021unsupervisedcocondenser,nie2020dc}. 
Currently, the dominant implementation is dense retrieval \cite{xiongdcn+mle,pipeline4}, which encodes the query and documents into dense embedding vectors to capture rich semantics.

\heading{\Acl{GR}} 
An emerging alternative to dense retrieval in document retrieval is \acfi{GR} \citep{DSI,tang-2023-recent}. 
It employs a sequence-to-sequence (Seq2Seq) architecture to generate relevant document identifiers (docids) for queries. 
In this manner, the knowledge of all documents in the corpus is encoded into the model parameters, similar to the human cognitive associative mechanism \cite{kounios2001cognitive,anderson2014human}.
To achieve this, \ac{GR} involves two basic operations \cite{DSI}:
\begin{enumerate*}[label=(\roman*)]
    \item \emph{indexing}, which memorizes the entire corpus by associating each document with its identifier, and 
    \item \emph{retrieval}, which uses the indexed corpus information to produce a ranked list of potentially relevant docids for a given query.
\end{enumerate*}

Using general language models, such as BART \cite{Lewis2019BARTDS} and T5 \cite{raffel2020exploringt5}, as the base Seq2Seq model has become a popular choice in GR \cite{zhuang2022bridgingdsiqg,genre,seal}. 
On top of this, some work has designed pre-training objectives for GR. 
For example, \citet{zhou2022ultron} proposed indexing- and retrieval-based pre-training tasks; document pieces or pseudo-queries are used as input, and docids (e.g., product quantization code) are predicted as output with maximum likelihood estimation (MLE). 
Similarly, \citet{chen2022corpusbrain} proposed retrieval-based tasks, which aim to construct and learn pairs of pseudo-queries and docids (i.e., Wikipedia titles) from the corpus.
These works demonstrate that applying specialized pre-trained models to GR yields superior results compared to using general language models.

\heading{Research challenges}
While pre-training methods have shown their effectiveness, important limitations remain in the following:
\begin{enumerate*}[label=(\roman*)]
    \item \textit{The construction process of pre-defined docids is independent from the pre-training process}. 
    This results in a semantic gap between both processes, which could potentially hinder the retrieval performance. 

    \item \textit{Docids remain unchanged during pre-training.} 
    If the initial docids are not suitable, they cannot be further adjusted after the training begins. 
    Consequently, it may become challenging to learn semantics and relationships between documents, impeding the achievement of satisfactory retrieval performance.

    \item \textit{Existing pre-training methods do not explicitly consider the interrelations between document-docid or query-docid pairs.} The widely-used MLE objective may result in difficulties in distinguishing among similar documents and docids. Therefore, we argue that the model should enhance its discriminative and generalization ability.
\end{enumerate*}

\heading{Approach}
To address these challenges, we introduce a general bootstrapped pre-training method for \ac{GR}, called BootRet.
Our objective is to dynamically adjust docids in accordance with the evolving model parameters during pre-training. 
The key idea is inspired by that the human brain updates the organization of existing knowledge to better match updated goals or contents in learning  \cite{mack2016dynamic}.      
BootRet includes three key steps:
\begin{enumerate*}[label=(\roman*)]
    \item \emph{Initial docid generation}. We leverage the encoder of the initial model to encode documents and then obtain the product quantization code \cite{zhan2021jointly,ge2013optimized} as the initial docids.

    \item \emph{Pre-training}. We design two pre-training tasks, i.e., corpus indexing task and relevance predication task. 
    The corpus indexing task aims to memorize corpus information and distinguish among similar documents and docids. We construct pairs of original documents and corresponding identifiers to simulate the indexing operation. 
    To enhance discrimination and generalization, we use a large language model (LLM) to generate noisy documents similar to the originals,  creating pairs of noisy documents and identifiers.  
    Besides, we design contrastive losses to help the model memorize and contrast these pairs.
    The relevance prediction task aims to learn relevance information from the corpus. 
    We construct pairs of  pseudo-queries and relevant docids to simulate the retrieval operation. 
    We also use a LLM to generate high-quality pseudo-queries for original documents as input and design a contrastive loss for the model to predict and contrast docids. 
    These two tasks are jointly learned, with the docids remaining fixed throughout this process.

    \item \emph{Enhanced bootstrapping}. The encoder of the model pre-trained with the above two tasks is further used to encode documents, updating document representations, and then updating the PQ code, i.e., docids. 
    These updated docids are further used to retrain the model based on the pre-training tasks. 
    
\end{enumerate*}
Steps (ii) and (iii) iteratively update the model parameters and docids.

We pre-train BootRet based on two kinds of large scale text corpus, i.e., MS MARCO \cite{msmarco} and Wikipedia \cite{wikidump}.
We then fine-tune BootRet on two representative downstream datasets widely used in GR research. 
The empirical experimental results show that BootRet can achieve significant improvements over strong GR baselines.

\heading{Contributions}
Our main contributions are:
\begin{enumerate*}[label=(\roman*)]
    \item  We propose a bootstrapped pre-training framework for GR to iteratively update the model parameters and docids. 
    \item BootRet demonstrates superior performance in downstream retrieval tasks. For instance, on the MS MARCO dataset, it outperforms the strong pre-training GR baseline, Ultron \cite{zhou2022ultron}, by 11.8\% in terms of Hits@1.
    \item Additionally, BootRet exhibits better zero-shot performance than other general language models.
\end{enumerate*}

%% file: sections/related_work.tex
\vspace{-2mm}
\section{Related Work}

\heading{Generative retrieval}
GR marks a new paradigm in document retrieval that generates identifier strings of documents as the retrieval target \cite{modelBased,DSI}. 
The current design of docids can be categorized into two types. 
\begin{enumerate*}[label=(\roman*)]
\item \textit{Pre-defined static docids.} They remain unchanged during training, such as document titles \cite{genre}, URLs \cite{ren2023tome}, product quantization code \cite{chen2023-continual,mehta2022dsi++}. 
This design is simple and shows decent performance \cite{chen2022corpusbrain}, but the pre-defined process is independent of training. 
\item \textit{Learnable docids.} They are optimized jointly with the retrieval task \cite{sun-2023-learning-arxiv,wang2023novo}. Though these docids are dynamic, their optimization primarily targets retrieval. 
Nevertheless, docids serve functions in both indexing and retrieval. 
\end{enumerate*}

In addition to widely-used supervised learning approaches \cite{sun-2023-learning-arxiv,zhang2023term-sets-arxiv,zhou-2023-enhancing}, recent studies \cite{zhou2022ultron,chen2022corpusbrain,zeng2023scalable} have explored pre-training for GR. 
However, each study adopts fixed docids, ignoring the potential mismatch between docids and the updated model. In contrast, our work dynamically updates both the docids and the evolving model to enhance the effectiveness.

\heading{Bootstrapping} The idea of bootstrapping training methods have garnered significant interest in various natural language processing tasks \cite{wu2009domain,deepika2021pattern,song2014dataless}. 
The approach involves generating new training data or information based on the previous model to iteratively enhance its capabilities. 
While the techniques are sometimes used in conjunction with supervised learning \cite{deepika2021pattern,song2014dataless}, our scenario involves an unlabeled corpus without ideal docids. 
Thus, we adopt a more unsupervised approach, iteratively refining the GR model and docids.

\heading{Dense retrieval} Dense retrieval \cite{karpukhin2020dense,gao2021unsupervisedcocondenser} is currently the de facto solution for document retrieval. 
It focuses on representing documents and queries as dense vectors in continuous spaces, capturing semantic relationships. 
Efficient vector search is facilitated by approximate nearest neighbor  \cite{xiong2020approximate} algorithms. 
Further enhancements include using pre-trained models within a dual-encoder architecture \cite{zhan2020repbert,nie2020dc} and hard negative mining techniques \cite{pipeline4,pipeline3}. 
Compared to dense retrieval, GR could achieve end-to-end global optimization.
However, its current performance lags behind state-of-the-art methods in dense retrieval.

%% file: sections/method.tex
\vspace{-4mm}
\section{Method}
\begin{figure*}[t]
    \centering
    \includegraphics[width=\textwidth]{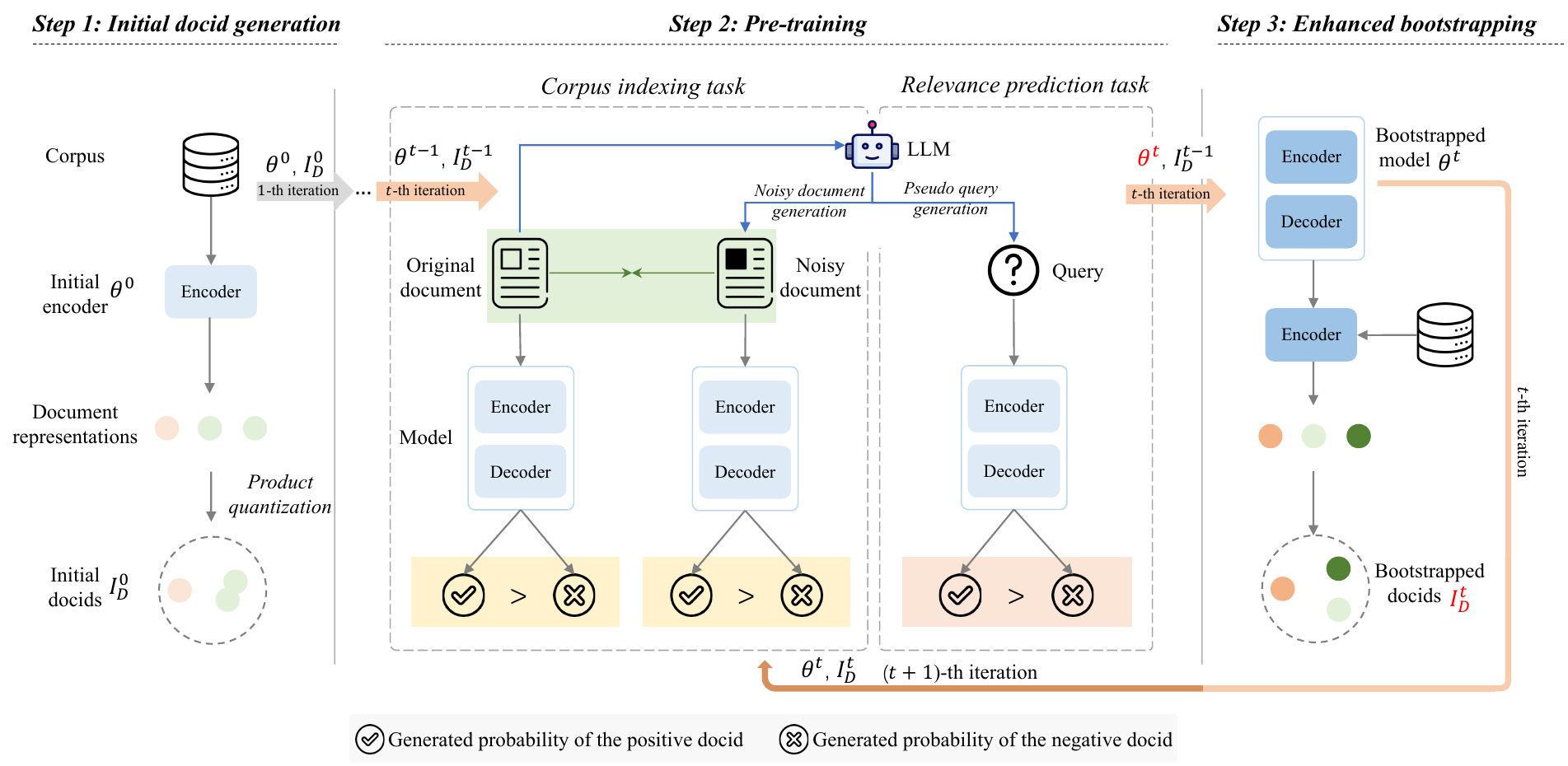}
    \caption{The bootstrapped pre-training pipeline of BootRet. (1) The initial docids $\mathcal{I}_\mathcal{D}^0$ are obtained with the initial model parameters $\theta^0$. (2) To perform the $t$-th iteration, we design the corpus indexing task and relevance prediction task for pre-training. We construct noisy documents and pseudo-queries with a LLM, and design contrastive losses (the yellow and the orange rectangles) and a semantic consistency loss (the green rectangle) to learn the corpus and relevance information discriminatively. After pre-training, the model updates from $\theta^{t-1}$ to $\theta^{t}$. (3) The bootstrapped $\theta^{t}$ is used to dynamically update the docids $\mathcal{I}_\mathcal{D}^{t-1}$ to $\mathcal{I}_\mathcal{D}^{t}$, i.e., bootstrapped docids, which are further used in the next iteration. (Figure should be viewed in color.)}
    \label{fig:overview}
    \vspace{-3mm}
\end{figure*}

\vspace{-2mm}
This section introduces the details of the BootRet model 
proposed in this paper. 
As shown in Figure \ref{fig:overview}, 
\begin{enumerate*}[label=(\roman*)]
    \item Given a corpus $\mathcal{D}=\{d_1,\ldots,d_{|\mathcal{D}|}\}$, we first construct an initial docid $id_i$ for each document $d_i$ in $\mathcal{D}$.
    The initial docid set is denoted as $\mathcal{I}_{\mathcal{D}}^0$.
    We employ an encoder-decoder language model as the base model, where initial parameters are denoted as $\theta^0$.
    \item Then, while keeping $\mathcal{I}_{\mathcal{D}}^0$ unchanged, we carefully design two pre-training tasks. During pre-training, the model parameters are updated from $\theta^0$ to $\theta^1$.
    \item Subsequently, fixing $\theta^1$, we update the docids to $\mathcal{I}_{\mathcal{D}}^1$, thus completing one iteration. 
    The updated docids can be used to further retrain the model for a next iteration.
\end{enumerate*}
We define the $t$-th iteration as updating the model parameters from $\theta^{t-1}$ to $\theta^t$ with fixed $\mathcal{I}_{\mathcal{D}}^{t-1}$ in Step (ii), and then based on fixed $\theta^t$, updating $\mathcal{I}_{\mathcal{D}}^{t-1}$ to $\mathcal{I}_{\mathcal{D}}^t$ in Step (iii).

\subsection{Model Architecture}
Like previous GR research \cite{DSI,NCI,chen2022corpusbrain}, we leverage a transformer-based model comprising: 
\begin{enumerate*}[label=(\roman*)]
\item An encoder, a bidirectional encoder to encode documents or pseudo-queries.
\item An identifier decoder, operating through a sequential generation process to produce document identifiers.
\end{enumerate*}
We initialize the model with T5-base \cite{raffel2020exploringt5}, and the initial model parameter is denoted as $\theta^0$.

\vspace{-2mm}
\subsection{Initial Docid Generation}\label{subsec:initial-docid-design}

Docids with semantic ties to the document content aid the model's learning \cite{DSI}. For effective bootstrapping, docids need efficient updates based on the model's progress. Considering these needs, we choose the widely used PQ code \cite{chen2023-continual} as the docid.

Specifically, we first encode all the documents to obtain document vectors with the encoder of $\theta^0$.
Following \cite{zhou2022ultron}, vectors are evenly divided into $g$ groups. For each group, we apply the $K$-means clustering algorithm to obtain $k$ cluster centers. Then, the docid can be represented by cluster indices of length $g$ corresponding to the clusters. 
And we denote the initial docid set as $\mathcal{I}_\mathcal{D}^0$.
To facilitate the generation of docids, we include all cluster indices from all groups obtained in the docid generation process as new tokens added to the model vocabulary.
$\mathcal{I}_\mathcal{D}^0$ will be used for subsequent iterative pre-training and updates. 

\subsection{Pre-training Tasks}
\label{subsec:pretraining-tasks}
The core idea is to construct pseudo document-docid pairs and query-docid pairs to simulate the indexing and retrieval operations, respectively. 
Our two pre-training tasks are: 
\begin{enumerate*}[label=(\roman*)]
    \item \textit{Corpus indexing task.} We first construct pairs of original documents and their corresponding identifiers. 
    For original documents, we use a LLM to construct similar but noisy documents $\mathcal{\Tilde{D}}$.
    $\Tilde{d}_i^h \in \mathcal{\Tilde{D}}$ is the $h$-th noisy version of $d_i$. 
    And we design multiple losses to guide the model to learn the associations between original or noisy documents and their identifiers.

    \item \textit{Relevance prediction task.} 
    We use a LLM to construct pseudo-queries $\mathcal{Q}$, and pair them with relevant docids.
    For $d_i$, we construct a total of $X$ queries, and $q_i^x \in \mathcal{Q}$ denotes the $x$-th pseudo-query.
\end{enumerate*}

\subsubsection{Corpus Indexing Task}

We introduce the construction of noisy documents and pre-training objectives in detail.

\heading{Noisy document construction}
The noisy documents should maintain semantic consistency with the originals while remaining distinguishable. 
We propose leveraging a LLM to effectively
achieve this. 
Inspired by \cite{raffel2020exploringt5}, we design the following four prompts to guide LLM generation: 
\begin{itemize}[leftmargin=*,nosep]
    \item A synonym replacement prompt: ``\texttt{Replace some words in the following document with their synonyms while maintaining the overall semantic meaning: \{d\}.}''
    \item A sentence removal prompt: ``\texttt{Remove one or more sentences from the following document, while maintaining the overall semantic meaning: \{d\}.}''
    \item A sentence shuffling prompt: ``\texttt{Rearrange the sentences in the following document to create a new flow, while maintaining the overall semantic meaning: \{d\}.}
    \item A word masking prompt: \texttt{Mask some words with [Masked] in the following document, while maintaining the overall semantic meaning: \{d\}.}''
\end{itemize}
Combining these prompts with an original document as the input, LLM generates four noisy documents, sharing the same docid with the original.

\heading{Pre-training objective}
In the $t$-th iteration, the objective consists of three parts as the follows.
\begin{itemize}[leftmargin=*,nosep]
\item \textbf{Semantic consistency loss}: It aims at maintaining overall semantic consistency between original and noisy documents. Specifically, in a mini-batch, there are a total of $4N$ document-docid pairs, where $N$ pairs correspond to the original pairs, and each original document has four noisy pairs.
This loss $ L_{SC} (\mathcal{D}, \mathcal{\Tilde{D}};\theta^{t-1})$ is defined as:
\begin{equation}\label{eq:semantic-consistency}
  \sum^N_{i=1}\sum^4_{h=1} 1- \operatorname{sim}(\operatorname{Enc}(d_i),\operatorname{Enc}(\Tilde{d}_i^h)),
\end{equation}
where  $\theta^{t-1}$ denotes model parameters of the previous iteration and $\operatorname{sim}(\cdot)$ is the cosine function.

\item \textbf{Contrastive losses for corpus indexing}: Conditioned on original document-docid pairs, we encourage the model to generate a docid that corresponds to the document rather than the docids of other documents. 
In the same mini-batch, we aim for the model to generate the docid corresponding to the document with a higher probability than generating others. Inspired by contrastive learning \cite{khosla2020supervised}, this loss $L_{C1} (\mathcal{D},\mathcal{I}_\mathcal{D};\theta^{t-1})$ is formalized as:
\begin{equation}
\label{eq:contrastive-original}
    - \sum^N_{i=1}\log \frac{\exp(P(id_i \mid d_i)/ \tau)}{\sum^{N}_{j=1}  \exp (P(id_j \mid d_i)/ \tau)},
\end{equation}
where $\tau$ is the temperature hyperparameter. $P(id_i \mid d_i)$ is the generated likelihood probability of $id_i$ conditioned on $d_i$. 
Similarly, for noisy pairs, the loss $L_{C2}(\mathcal{\Tilde{D}},\mathcal{I}_\mathcal{D};\theta^{t-1})$ is:
\begin{equation}
\label{eq:contrastive-noise}
     -\sum^N_{i=1}\sum^4_{h=1} \log \frac{\exp(P(id_i \mid \Tilde{d}_i^h)/ \tau)}{\sum^{N}_{j=1}  \exp (P(id_j \mid \Tilde{d}_i^h) / \tau)}.
\end{equation}
\end{itemize}
The pre-training objective of the corpus indexing task $L_{CI} (\mathcal{D}, \mathcal{\Tilde{D}},\mathcal{I}_\mathcal{D};\theta^{t-1})$ is a weighted sum of the three aforementioned losses, denoted as:
\begin{equation}\label{eq:pre-train-corpus}
     L_{SC}(\cdot) + \alpha L_{C1}(\cdot) + \beta L_{C2}(\cdot),
\end{equation}
where $\alpha$ and $\beta$ are hyperparameters.

\subsubsection{Relevance Prediction Task}
We introduce the construction process of pseudo-queries, and the pre-training objective.

\heading{Pseudo-query construction}
To generate high-quality pseudo-queries for the original documents, we employ a LLM using the prompt: ``\texttt{Given the following document \{d\}, generate  \{X\} insightful queries that a reader might have after reading the content. Ensure the queries cover key concepts.}''
When the prompt is combined with a document $d_i$ and the required number of pseudo-queries $X$ as input, we obtain well-written pseudo-queries.
They share the same docids as the input original document.

\heading{Pre-training objective}
Similarly, we ensure that 
the model tends to generate relevant docids than irrelevant ones.
In the same mini-batch, 
the loss $L_{RP} (\mathcal{Q},\mathcal{I}_{\mathcal{Q}};\theta^{t-1})$ in the $t$-th iteration is:
\begin{equation}\label{eq:pre-train-retrieval}
      -\sum^N_{i=1} \sum^X_{x=1} \log \frac{\exp(P(id_i \mid q^x_i)/ \tau)}{\sum^{N}_{j=1}  \exp (P(id_j \mid q^x_i) / \tau)}.
\end{equation}

\subsubsection{Joint Learning}
We jointly pre-train the model with two above objectives and two sequence generation objectives. 
In the $t$-th iteration, the overall loss $L_{Pre}(\mathcal{D},\mathcal{\Tilde{D}},\mathcal{I}_\mathcal{D},\mathcal{Q},\mathcal{I}_\mathcal{Q};\theta^{t-1})$ is :
\begin{equation}\label{eq:pre-train}
   \gamma L_{CI}(\cdot) + \rho L_{RP}(\cdot) +  \lambda L_{ID}(\cdot) +  \lambda L_{RE}(\cdot), 
\end{equation}
where $\gamma$, $\rho$ and $\lambda$ are hyperparameters; 
$L_\mathit{ID}(\mathcal{D},\mathcal{\Tilde{D}},\mathcal{I}_\mathcal{D};\theta^{t-1})$ is the widely used standard MLE loss based on document-docid pairs: 
\begin{equation}
\mbox{}\hspace*{-2mm}
      - \sum^{|\mathcal{D}|}_{i=1} \log P(id_i \!\mid\! d_i) \!-\! \sum^{|\mathcal{D}|}_{i=1} \sum^4_{h=1}\log P(id_i\!\mid\! \Tilde{d}_i^h).
\hspace*{-1mm}\mbox{}
     \label{eq:indexing} 
\end{equation}
Note, Eq.~\eqref{eq:contrastive-original} and Eq.~\eqref{eq:contrastive-noise} ensure that the model's probability of generating the corresponding docid is greater than generating other docids. Eq.~\eqref{eq:indexing} does not explicitly contrast with other docids.
$L_\mathit{RE}(\mathcal{Q},\mathcal{I}_\mathcal{D};\theta^{t-1})$ is based on query-docid pairs:
\begin{equation}
      - \sum^{|Q|}_{i=1} \log P(id_i\mid q_i).
\label{eq:retrieval} 
\end{equation}
%
%
During training, we construct two types of batch data. One type has original and noisy documents-docid pairs, optimized using Eq.~\eqref{eq:pre-train-corpus} and Eq.~\eqref{eq:indexing}.
The other type has pairs of the pseudo-query and relevant docid only, optimized using  Eq.~\eqref{eq:pre-train-retrieval} and Eq.~\eqref{eq:retrieval}. 
After jointly training during the $t$-th iteration, $\theta^{t-1}$ updates to $\theta^{t}$ with docids fixed.

\subsection{Enhanced Bootstrapping Strategy} 
\label{subsec:bootstrapping}
Based on the updated $\theta^{t}$, we introduce how to update docids $\mathcal{I}_\mathcal{D}^{t-1}$ and retrain the model.

\heading{Docid update}
Fixing $\theta^{t}$, we use the encoder of $\theta^{t}$ to encode documents as in Section \ref{subsec:initial-docid-design}, to update docids of the previous iteration $\mathcal{I}_\mathcal{D}^{t-1}$, to $\mathcal{I}_\mathcal{D}^{t}$.
We refer to the version following the initial iteration's completion (i.e., $\mathcal{I}_\mathcal{D}^{1}$ and $\theta^{1}$) as \textbf{BootRet-Bs}.

\heading{Retrain the model}
To proceed to the next iteration, we retrain the model with $\mathcal{I}_\mathcal{D}^{t}$ as described in Section \ref{subsec:pretraining-tasks}.
After multiple iterations, we achieve continuous dynamic alignment and enhancement. We refer to this version as \textbf{BootRet-Mt}.

%% file: sections/experimental_settings.tex
\section{Experimental Settings}

\heading{Pre-training corpus} 
For pre-training we use two large, publicly available corpora:
\begin{enumerate*}[label=(\roman*)]
    \item \emph{English Wikipedia}, which contains tens of millions of well-written documents and we downloaded this dump \cite{wikidump} for pre-training, and 
    \item the \emph{MS MARCO Document Collection} \cite{msmarco}, which has about 3 million documents extracted from web documents using the Bing search engine.
\end{enumerate*}
For each corpus, we sample 500K documents and generated four noisy documents and five pseudo-queries, i.e., $X$, for each document. This results in 2.5M documents and 2.5M pseudo-queries for pre-training.
BootRet-BS$^\mathit{Wiki}$ and BootRet-BS$^\mathit{MS}$ denote the model pre-trained on Wikipedia and MS MARCO, respectively.

\heading{Downstream retrieval datasets} We leverage two representative retrieval datasets.
\begin{enumerate*}[label=(\roman*)]
    \item \emph{MS MARCO Document Ranking dataset} \cite{msmarco}. Following the setup of \cite{zhou2022ultron}, we sample a subset of 300K documents for experimentation, denoted as MS 300K, containing 360K training queries, 6980 evaluation queries.
    These documents do not overlap with the ones used in pre-training.

    \item \emph{Natural Question} (NQ) \cite{naturalquestion} has real questions and Wikipedia documents, having about 228K documents with 307K training queries and 7.8K test queries.
\end{enumerate*}

\heading{Baselines}
Following typical GR research \cite{DSI,NCI}, we examine three types of baseline:
\begin{enumerate*}[label=(\roman*)]
    \item \emph{Sparse retrieval baselines}: BM25 \cite{bm25}, and DocT5Query \cite{doct5query}.

    \item \emph{Dense retrieval baselines}: RepBERT \cite{zhan2020repbert}, DPR \cite{karpukhin2020dense}, and ANCE \cite{xiong2020approximate}. 

    \item \emph{Advanced GR baselines}: DSI \cite{DSI}, GENRE \cite{genre}, SEAL \cite{seal}, DSI-QG \cite{zhuang2022bridgingdsiqg}, NCI \cite{NCI}, Ultron-PQ \cite{zhou2022ultron}, Corpusbrain \cite{chen2022corpusbrain},   GenRet \cite{sun-2023-learning-arxiv}, and NOVO \cite{wang2023novo} . 
\end{enumerate*}
For more details on our baselines, please refer to Appendix~\ref{appendix-baselines}.

\heading{Evaluation metrics} 
Following GR work \cite{DSI,li2023multiview}, for NQ, we use hit ratio ({Hits$@K$}) with $K=\{1,10\}$ as the metric. 
For MS 300K, we also use mean reciprocal rank ({MRR$@K$}) with $K=\{3,20\}$ \cite{li2023multiview}.

\heading{Implementation details} For pre-training, we use the LLaMA-13b model \cite{LLaMA13b} to generate noisy documents and pseudo-queries. We initialize our model with T5-base (220M) \cite{t5base}. For docids, we set the length $g$ of PQ codes to 24, the number of clusters $k$ to 256, and the dimension of vectors to 768. The hyperparameters for pre-training are set to $\alpha=\beta=\lambda=1$, $\gamma=\rho=2$ and $\tau=0.2$. The batch size is 256. The Adam optimizer with a learning rate of 5e-5 is used, and the sequence length of documents is set to 512.
The max training step is 500K, with the first iteration occurring at step 100K, followed by iterations every 40K steps thereafter.

For fine-tuning, 
we use the pre-trained model obtained from the last iteration to generate docids. 
Models are further fine-tuned with document-docid pairs and labeled query-docid pairs with MLE \cite{DSI}. 
Following \cite{NCI,zhou2022ultron}, we additionally generate 10 pseudo-queries for each document to enhance training. 
We set the learning rate as 1e-3, and the max training steps as 30K. Other settings remain consistent with the pre-training stage.

All models are trained on eight NVIDIA Tesla A100 80GB GPUs.
For inference, we build a prefix trie \cite{genre} for docids and use constrained beam search with 20 beams to decode docids. 
For more details, please see Appendix~\ref{appendix-implementation}.

%% file: sections/experimental_resutlts.tex
\vspace{-2mm}
\section{Experimental Results}
This section presents the experimental findings. 

\begin{table}[t]
    \centering
    \setlength{\tabcolsep}{3.5pt}
    \renewcommand{\arraystretch}{0.85}
    \begin{tabular}{l ccccc ccccc}
        \toprule
        \multirow{2}{*}{\textbf{Method}} &  
        \multicolumn{2}{c}{\textbf{MRR}} & 
        \multicolumn{2}{c}{\textbf{Hits}}\\
        \cmidrule(r){2-3}
        \cmidrule(r){4-5}

        & \textbf{$@3$} & \textbf{$@20$} & \textbf{\phantom{1}$@1$\phantom{1}} & \textbf{$@10$}   \\
        \midrule 

        BM25 &  22.57 & 26.67 & 24.78 & 40.73\\
        DocT5query & 27.38 & 29.63 & 30.13 & 46.93\\

        \midrule
         RepBERT & 31.47 & 33.68 & 33.16 & 55.83\\
         DPR & 34.84 &	36.79 &	36.52 &	58.68  \\
         ANCE & 30.76 & 34.25 & 33.63 & 53.62\\
        \midrule
        	 
         DSI & 23.21 &	28.93 	&28.14 	&49.72\\
         GENRE &31.12 &	33.49 	&33.18 	&53.56 \\
         SEAL   & 31.35 	&33.57 &	33.34 &	53.74\\
         DSI-QG & 33.64 &	35.81 &	34.96 &	58.62\\
         NCI  & 33.86 	&36.20 	&35.02 &	59.21\\
         Corpusbrain & 34.72 &37.25  &	36.14  &60.32\\
         Ultron-PQ & 35.25 	 &38.41  &	39.53  &	62.85\\

         GenRet&	37.26 &	40.53 &	41.68 &	64.92 \\
         NOVO&	38.36 &	41.29 &	43.14 &	64.55 \\

         \midrule
         BootRet-Bs$^\mathit{Wiki}$& 36.28\rlap{$^{\ast}$} &	39.25\rlap{$^{\ast}$} &	40.73\rlap{$^{\ast}$} &	63.78\rlap{$^{\ast}$}\\
         BootRet-Bs$^\mathit{MS}$& 37.13\rlap{$^{\ast}$} &	40.48\rlap{$^{\ast}$} 	&41.56\rlap{$^{\ast}$} 	&64.89\rlap{$^{\ast}$}\\
         BootRet-Mt$^\mathit{Wiki}$& 38.83\rlap{$^{\ast}$} 	&41.36\rlap{$^{\ast}$} &	43.97\rlap{$^{\ast}$} &	65.83\rlap{$^{\ast}$}\\
         BootRet-Mt$^\mathit{MS}$& \textbf{39.35}\rlap{$^{\ast}$} &	\textbf{42.79}\rlap{$^{\ast}$} &	\textbf{44.21}\rlap{$^{\ast}$} &	\textbf{66.73}\rlap{$^{\ast}$} \\

        \bottomrule
    \end{tabular}
    \caption{Retrieval performance on MS 300K. The best results are shown in \textbf{bold}.  $\ast$ indicates statistically significant improvements over the best performing GR baseline NOVO  ($p \leq 0.05$).}
    \label{tab:main-results-ms}
    \vspace*{-2mm}
\end{table}

\subsection{Main Results}
The comparison between our BootRet and baselines on MS 300K and NQ
are shown in Table~\ref{tab:main-results-ms} and Table~\ref{tab:main-results-nq}, respectively.
We observe:
\begin{enumerate*}[label=(\roman*)]
\item Dense retrieval baselines generally outperform sparse retrieval baselines, indicating that dense vectors capturing rich semantics are more beneficial for retrieval.

\item Dense retrieval baselines outperform naive GR methods, such as DSI and SEAL, demonstrating the challenge of learning with only labeled data for GR.

\item  DSI-QG and NCI with data augmentation perform better than dense retrieval baselines, suggesting that GR requires more labeled data.

\item Pre-trained baselines, i.e., Ultron and Corpusbrain, outperform supervised learning GR baselines, highlighting the necessity of pre-training for GR.

\item Our BootRet-Mt$^\mathit{Wiki}$ and BootRet-Mt$^\mathit{MS}$ outperform base versions and Ultron, demonstrating the effectiveness of bootstrapped pre-training with dynamic identifiers.

\item In MS300K, our BootRet-Bs does indeed perform slightly worse compared to strong GR baselines such as UniGen, GenRRL, GenRet, NOVO, and ASI. However, the performance of BootRet-Mt is better than them, which also demonstrates the effectiveness of our approach. Similar conclusions are observed in NQ.

\item BootRet-Bs$^\mathit{MS}$ performs better on MS 300K than BootRet-Bs$^\mathit{Wiki}$, while the opposite is observed on NQ, indicating that the performance of pre-trained models improves when downstream data and pre-training corpora are more similar.
\end{enumerate*}
Additionally, Table~\ref{appendix-full-ranking} in the appendix, shows that the performance of GR methods lags  behind cross-encoder methods, suggesting ample room for exploration in GR.

\begin{table}[t]
    \centering
    \setlength{\tabcolsep}{11.5pt}
    \renewcommand{\arraystretch}{0.85}
    \begin{tabular}{l ccccc ccccc}
        \toprule
        \textbf{Method} &  
        \textbf{Hits$@1$} &
        \textbf{Hits$@10$}\\
        \midrule
        BM25\rlap{$^\star$} & 29.27 & 60.16 \\
        DocT5query\rlap{$^\star$} & 39.13 & 69.72\\

        \midrule
        RepBERT & 50.20 & 78.12\\
        DPR\rlap{$^\star$} & 52.63 & 79.31\\
        ANCE & 45.42 & 72.75\\
        \midrule

         DSI\rlap{$^\star$} & 27.40 & 56.60\\
         GENRE\rlap{$^\star$}  & 26.30 & 71.20\\
         SEAL\rlap{$^\star$}   & 26.30 & 74.50 \\
         DSI-QG\rlap{$^\star$} & 63.49 & 82.36\\
         NCI  & 64.24 & 83.11\\
         Corpusbrain &65.12 	&84.09 \\
         Ultron-PQ &64.61 &	84.45  \\

        GenRet &65.42 &	85.67\\
        NOVO & 66.13 	&86.24\\

         \midrule
         BootRet-Bs$^\mathit{Wiki}$& 66.71\rlap{$^{\ast}$} &	85.53\rlap{$^{\ast}$}\\
         BootRet-Bs$^\mathit{MS}$&65.88 &	85.04 \\
         BootRet-Mt$^\mathit{Wiki}$& \textbf{67.32}\rlap{$^{\ast}$} 	&\textbf{87.59}\rlap{$^{\ast}$} \\
         BootRet-Mt$^\mathit{MS}$& 66.15\rlap{$^{\ast}$} 	&86.31\rlap{$^{\ast}$} \\
  
        \bottomrule
    \end{tabular}
    \caption{Retrieval performance on NQ. Methods marked with $\star$ indicate results are obtained from \cite{seal,DSI,zhuang2022bridgingdsiqg}. The best results are shown in \textbf{bold}. $\ast$ indicates statistically significant improvements over the best performing GR baseline NOVO  ($p \leq 0.05$).}
    \label{tab:main-results-nq}
    \vspace*{-2mm}
\end{table}

\begin{table}[t]
    \centering
    \setlength{\tabcolsep}{2mm}
    \renewcommand{\arraystretch}{0.85}
    \begin{tabular}{l @{} cc}
        \toprule
        \multirow{2}{*}{\textbf{Method}} &  \multicolumn{1}{c}{\textbf{MS 300K}} &  \multicolumn{1}{c}{\textbf{NQ}}\\
        \cmidrule(r){2-2}
        \cmidrule{3-3}

        & \textbf{Hits$@10$}  &  \textbf{Hits$@10$}  \\
        \midrule
        BootRet-Mt$^\mathit{Wiki}$ &65.83 &	87.59 \\
        \midrule
        \hspace{1mm} w/o dynamic identifiers& 63.14&	83.81 \\
        \midrule

        BootRet-Bs$^\mathit{Wiki}$& 63.78 &	85.53 \\
        \midrule
        \hspace{1mm} w/o pre-training  &59.95	 &83.26\\
        \hspace{1mm} w/o retrieval prediction  &63.01 &	83.82 \\
        \hspace{1mm} w/o corpus indexing & 63.28	 &83.91\\
        \hspace{1mm} w/o noisy documents & 63.47	 &84.17\\
         \hspace{1mm} w/o contrastive losses & 63.31		 &83.94\\
        \bottomrule
    \end{tabular}
    \caption{Ablation study of the pre-training components.}
    \label{tab:ablation-wiki}
    \vspace*{-4mm}
\end{table}

\subsection{Ablation Study}
To analyze the impact of each part of BootRet, we conduct ablation study on the Wikipedia pre-training corpus.
From Table~\ref{tab:ablation-wiki}, we observe the following:
\begin{enumerate*}[label=(\roman*)]
    \item When not using dynamic identifiers  (i.e., the 2nd row), wherein the model solely undergoes repeated pre-training using fixed docids, the performance significantly deteriorates compared to BootRet-Mt$^\mathit{Wiki}$, affirming the effectiveness of dynamic identifiers.
    
    \item When pre-training is not performed, and docids are directly obtained using the initial T5-base model (i.e., the 4th row), the model's performance is lower than that of Ultron. This underscores the necessity of pre-training for GR.
    
    \item When pre-training does not involve the retrieval prediction task or corpus indexing task (i.e., the 5th-6th rows), the performance is lower than BootRet-Bs$^\mathit{Wiki}$ (i.e., the 3rd row). This confirms that pre-training should consider both relevance and corpus information.
    
    \item Not learning the corpus indexing task (i.e., the 6th row) leads to better performance compared to Ultron, indicating that the contrastive loss in the retrieval prediction task enhances the discriminative ability.

    \item When the corpus indexing task does not use noisy documents (i.e., the 7th row), the performance is even lower, demonstrating that both noisy documents and contrastive losses contribute to discriminating similar documents and docids.

    \item When not using contrastive losses, i.e., the 8th row, where pre-training solely uses MLE losses (Eq.~\eqref{eq:indexing} and \eqref{eq:retrieval}) and Eq.~\eqref{eq:semantic-consistency}, there is a significant decrease in performance compared to BootRet-Bs$^\mathit{Wiki}$, indicating the effectiveness of contrastive losses.

\end{enumerate*}
The ablation results based on the MS MARCO pre-training corpus show similar trends, as shown in Table \ref{tab:ablation-ms}.

\begin{table}[t]
    \centering
    \setlength{\tabcolsep}{6.5pt}
    \renewcommand{\arraystretch}{0.85}
    \begin{tabular}{l ccccc ccccc}
        \toprule
        \multirow{2}{*}{\textbf{Methods}} &  \multicolumn{1}{c}{\textbf{MS 300K}} &  \multicolumn{1}{c}{\textbf{NQ}}\\
        \cmidrule(r){2-2}
        \cmidrule{3-3}

        & \textbf{Hits$@10$}  &  \textbf{Hits$@10$}  \\
        \midrule
        BootRet-Mt$^\mathit{MS}$& 66.73 & 	86.31  \\
        \midrule
        \hspace{1mm} w/o dynamic identifiers& 63.55& 	84.62 \\
        \midrule

        BootRet-Bs$^{MS}$&64.89 	&85.04 \\
        \midrule
       \hspace{1mm} w/o pre-training&  59.57& 83.71 \\
       \hspace{1mm} w/o retrieval prediction &  63.02	& 84.51\\
       \hspace{1mm} w/o corpus indexing &63.46	& 84.76 \\
       \hspace{1mm} w/o noisy documents& 63.95	& 84.96 \\
       \hspace{1mm} w/o contrastive  losses& 63.24	&84.62 \\
       \bottomrule
       \end{tabular}
      \caption{Ablation study of the pre-training components based on the MS MARCO pre-training corpus.}
    \label{tab:ablation-ms}
    \vspace{-2mm}
\end{table}

\begin{figure}[t]
     \vspace{-2mm}
     \centering
     \includegraphics[width=0.5\textwidth]{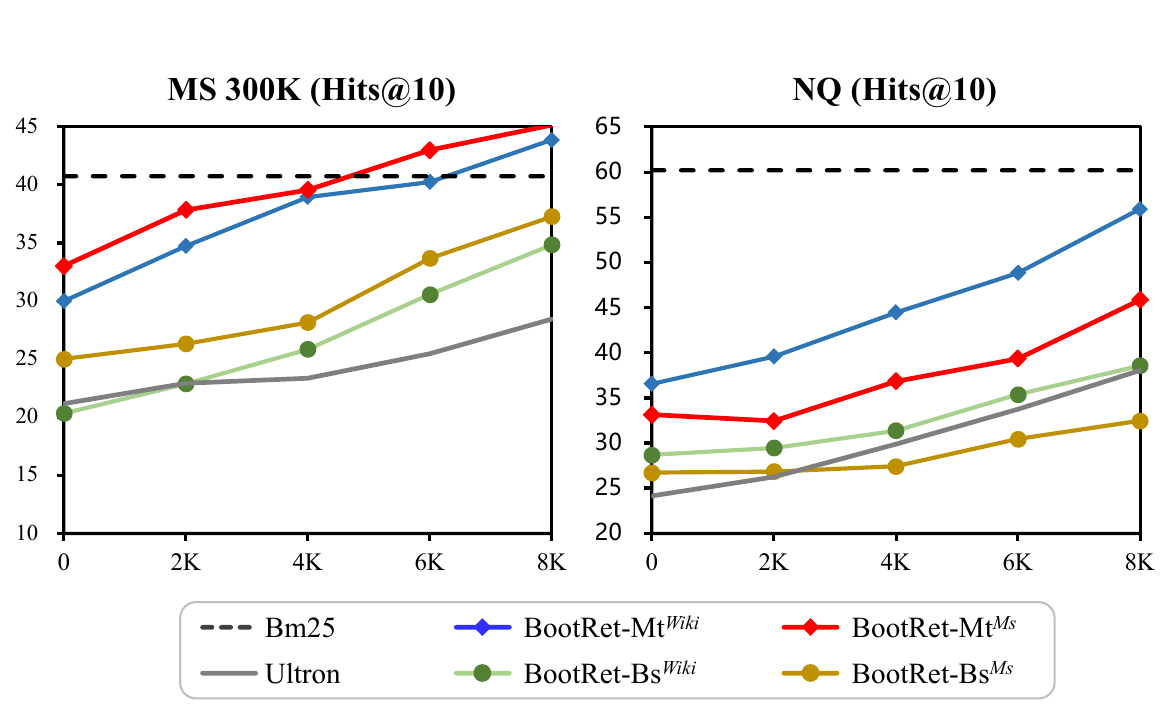}
     \caption{Results under zero- and low-resource setting. The x-axis indicates the number of labeled queries.}
     \label{fig:zero}
     \vspace{-2mm}
\end{figure}

\subsection{Zero- and Low-resource Settings}
To show whether BootRet can perform well with limited data, we randomly sample 2K, 4K, 6K, and 8K queries from the training set of both datasets.
From Figure \ref{fig:zero}, we observe the following:
\begin{enumerate*}[label=(\roman*)]
\item Under the zero-shot setting, where the model learns solely from the corpus without annotated queries, BootRet-Bs$^\mathit{Wiki}$ initially performs worse than Ultron on MS 300K. However, as fine-tuning with annotated queries progresses, BootRet-Bs$^\mathit{Wiki}$ quickly surpasses Ultron. This is possibly due to Ultron directly pre-training on the downstream dataset's corpus, while BootRet-Bs$^\mathit{Wiki}$'s pre-training corpus differs significantly from MS 300K. It also indicates that BootRet-Bs$^\mathit{Wiki}$ requires less annotated data to achieve rapid performance improvement.

\item Under the low-resource setting, both base versions of BootRet exhibit performance gaps compared to BM25, highlighting the importance of annotated data for GR.

\item Both versions of BootRet-Mt demonstrate better performance over base versions. Additionally, they achieve performance comparable to BM25 at approximately 1.3\%, i.e., 5K, queries fine-tuning on MS 300K.
\end{enumerate*}
Similar trends are observed for all methods on NQ, but all GR models perform worse than BM25.

\vspace{-2mm}
\subsection{Impact of the Number of Iteration}
\begin{figure}[t]
     \centering
     \includegraphics[width=0.5\textwidth]{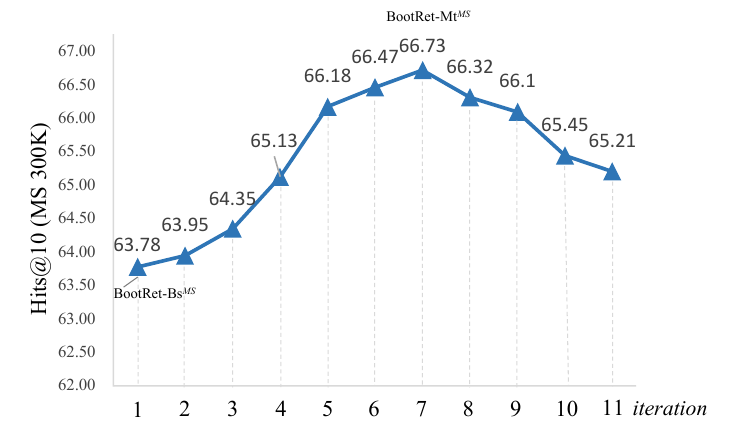}
     \caption{Retrieval performance of different number of iterations on MS 300K.}
     \label{fig:iteration}
\end{figure}

The iteration of updating docids and model parameters is important in our proposed bootstrapping pre-training method. 
We analyze the retrieval performance of the number of iterations on the downstream task, MS 300K,  pre-training on the MS MARCO corpus. 
In Figure~\ref{fig:iteration}, we find that performance generally improves as the number of iterations increases from 1 to 7, indicating the effectiveness of the bootstrapping pre-training method. However, performance begins to decline gradually after exceeding 7 iterations, possibly due to the model overfitting to the pre-training data.

\begin{figure}[t]
     \centering
     \includegraphics[width=0.5\textwidth]{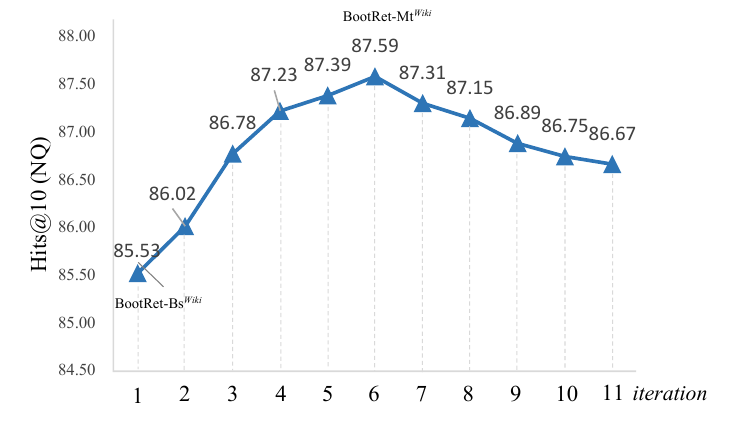}
     \caption{Retrieval performance of different number of iterations on NQ.}
     \label{fig:iteration-nq}
     \vspace{-4mm}
\end{figure}

As shown in Figure~\ref{fig:iteration-nq}, it show the retrieval performance of different number of iterations on NQ, which aligns with the trend on MS 300K. 
The phenomenon that the performance degrades substantially after a certain iteration, is reasonable. Because different datasets have different characteristics and properties. Therefore, the optimal number of iterations may vary. However, the optimal iteration range is similar across datasets. For example, we found that the best performance is achieved around the 7th iteration in MS 300K, while it is around the 6th iteration in NQ (Figure~\ref{fig:iteration-nq}). Therefore, for computational efficiency, when generalizing to other datasets, one can initially choose the number of iterations within a similar range.

\vspace{-2mm}
\subsection{Impact of Noisy Documents}
\begin{figure}[t]
     \vspace{-2mm}
     \centering
     \includegraphics[width=0.5\textwidth]{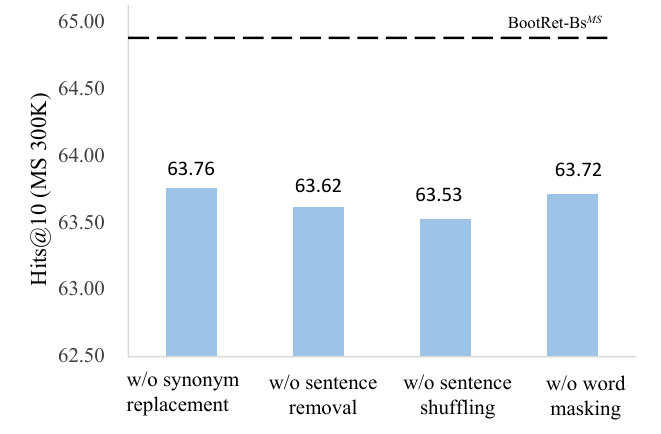}
     \vspace{-4mm}
     \caption{Retrieval performance of different prompts for generating noisy documents during pre-training.}
     \label{fig:ablation-noise-strategy}
     \vspace{-3mm}
\end{figure}

To analyze the impact of different prompts for generating noisy documents, we remove noisy documents generated using a certain type of prompt during pre-training to train BootRet-Bs$^\mathit{MS}$ and evaluate its retrieval performance on the downstream MS 300K dataset.
Based on Figure~\ref{fig:ablation-noise-strategy}, we observe the following.
\begin{enumerate*}[label=(\roman*)]
    \item When the noisy documents generated by the shuffling prompt are removed, the performance dropped the most (the red circle), likely due to the significant semantic differences introduced by altering sentence order, reinforced by semantic consistency loss, improving discrimination ability. The sentence removal prompt (the green circle) shows a similar result.
    
    \item Next, the word masking prompt (the grey circle) yields moderate results, possibly due to the omission of masked token prediction (as the initial T5 is already pre-trained for this task), thereby weakening the masking effect. 
    
    \item Lastly, the synonym replacement prompt (the blue circle) performs the most modestly, possibly because it introduces minimal semantic changes, thus having the same effect as original documents.
\end{enumerate*}

\begin{figure}[t]
     \centering
     \includegraphics[width=0.47\textwidth]{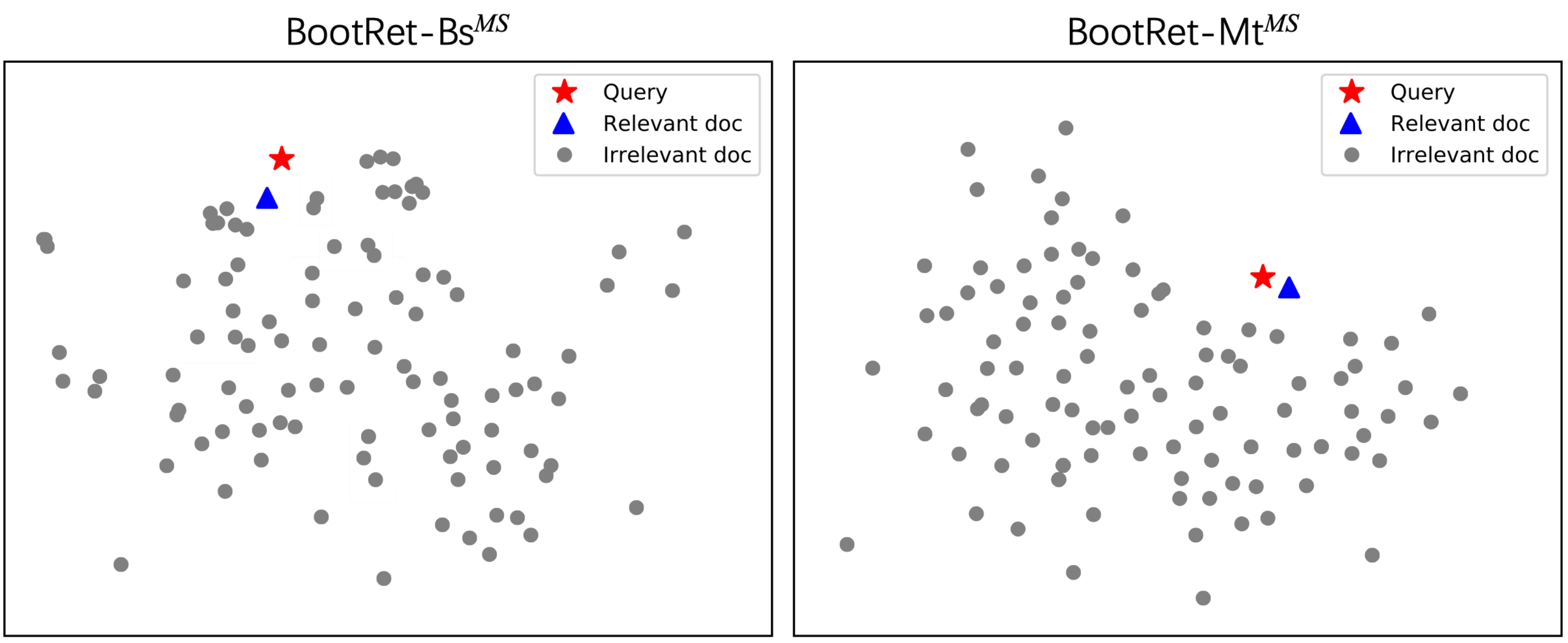}
     \caption{t-SNE plot of representations of a query (QID:1039861) from MS 300K validation set and documents corresponding to the generated  top-100 docid list by BootRet-Bs$^\mathit{MS}$ and BootRet-Mt$^\mathit{MS}$. }
     \label{fig:visual}
     \vspace{-5mm}
\end{figure}

\subsection{Visual Analysis}
To further analyze the bootstrapped pre-training, we conduct visual analysis on BootRet-Bs$^\mathit{MS}$ and BootRet-Mt$^\mathit{MS}$ on MS 300K. We sample a query, `` germany gasoline cost'' (QID: 194592), from the validation set and visualize the documents corresponding to the decoded docid lists (top 100) generated by BootRet-Bs$^\mathit{MS}$ and BootRet-Mt$^\mathit{MS}$. Specifically, we visualize  the query and document representations encoded by the encoders of both models.

From Figure~\ref{fig:visual}, we observe that compared to BootRet-Bs$^\mathit{MS}$ (left), BootRet-Mt$^\mathit{MS}$ (right) exhibits the relevant docid (the blue triangle) closer to the query (the red star), while irrelevant documents (the grey circles) are farther away.  
Additionally, we observe that irrelevant documents near the query are more clustered in BootRet-Bs$^\mathit{MS}$ compared to BootRet-Mt$^\mathit{MS}$, indicating that dynamic identifiers and pre-training tasks could effectively distinguish between documents.

%% file: sections/conclusion.tex
\section{Conclusion}
\vspace{-2mm}
In this work, we proposed BootRet, a bootstrapped pre-training method for GR, addressing the mismatch between pre-defined fixed docids and evolving model parameters in existing pre-training approaches. 
It dynamically adjusts docids based on the model pre-trained with two tasks.
Extensive experiments validate that BootRet achieves superior performance compared to strong GR baselines on downstream tasks, even in the zero-shot setting. 

\section*{Acknowledgements}
    This work was funded by the Strategic Priority Research Program of the CAS under Grants No. XDB0680102, the National Key Research and Development Program of China under Grants No. 2023YFA1011602 and 2021QY1701, the National Natural Science Foundation of China (NSFC) under Grants No. 62372431, the Youth Innovation Promotion Association CAS under Grants No. 2021100, the Lenovo-CAS Joint Lab Youth Scientist Project, and the project under Grants No. JCKY2022130C039. This work was also (partially) funded by the Hybrid Intelligence Center, a 10-year program funded by the Dutch Ministry of Education, Culture and Science through the Netherlands Organisation for Scientific Research, https://hybrid-intelligence-centre.nl, project LESSEN with project number NWA.1389.20.183 of the research program NWA ORC 2020/21, which is (partly) financed by the Dutch Research Council (NWO), and the FINDHR (Fairness and Intersectional NonDiscrimination in Human Recommendation) project that received funding from the European Union’s Horizon Europe research and innovation program under grant agreement No 101070212. All content represents the opinion of the authors, which is not necessarily shared or endorsed by their respective employers and/or sponsors.

%% file: sections/limitations.tex
\section{Limitations}

While BootRet has shown certain results in GR, it still has several limitations. 
\begin{enumerate*}[label=(\roman*)]

    \item In the relevance prediction task, although we incorporate negative samples, the computational cost limits our ability to conduct comprehensive comparisons beyond batch-level contrasts. Future work could explore integrating dynamic hard negative mining techniques from traditional retrieval methods into GR.

    \item We design prompts with minimal hyperparameters to generate noisy documents. Future research could explore corresponding hyperparameter designs, such as determining the extent of sentence shuffling/removal strategies.

    \item Compared to other GR pre-training methods, our pre-training incurs slightly higher computational costs due to the need to update docids at each iteration. In future work, we can further explore how to trade off iteration costs and performance.

    \item For handling incremental documents, we ignore this issue in this work. For future work, inspired by \cite{chen2023-continual}, we could adaptively adjust cluster centers in the docid generation process based on the similarity between new and old documents.

    \item Scalability is a significant challenge in current GR, requiring targeted solutions. Currently, a few works \cite{zeng2023scalable,pradeep2023does-scale-genir} are exploring this issue. Differently, our work focuses on pre-training for GR which can provide suitable base model for GR. Therefore, the size of our experimental datasets follows that of most current GR works \cite{sun-2023-learning-arxiv,wang2023novo,NCI,li2023multiview,DSI,zhou2022ultron}. We leave the scalability issue in the future. 

    \item \citet{Jin2023LanguageMA} is concurrent work with ours, proposing to conduct language model indexer pretraining and Docid learning jointly. We do not consider this in the present study.

\end{enumerate*}

%% file: sections/appendix.tex
\section{Appendix}

\subsection{Baseline Details}\label{appendix-baselines}
The baseline methods are described as follows:

\textit{Sparse retrieval baselines}:
\begin{enumerate*}[label=(\roman*)]
    \item BM25 \cite{bm25} is a widely used strong term-based method. We implement it based on the Anserini toolkit \cite{Anserini}; 
    \item DocT5Query \cite{doct5query} expands a document with pseudo-queries predicted by a fine-tuned T5 \cite{raffel2020exploringt5} conditioned on the original document. And then we perform the BM25 retrieval. 
 \end{enumerate*}

\textit{Dense retrieval baselines} :
\begin{enumerate*}[label=(\roman*)]
    \item DPR \cite{karpukhin2020dense} is a BERT-based dual-encoder model using dense embeddings for texts; 
    \item ANCE \cite{xiong2020approximate} leverages ANN algorithm and  hard negative techniques  for training a  dual-encoder model; and 
    \item RepBERT \cite{zhan2020repbert} is also a dual-encoder model with brute force searching. 
\end{enumerate*}

\textit{Advanced GR baselines}:
\begin{enumerate*}[label=(\roman*)]
    \item DSI \cite{DSI} employs semantic structured numbers as docids via a hierarchical k-means clustering algorithm. 
    \item GENRE \cite{genre} uses document titles as docids. It learns the document-docid pairs. 
    For NQ, it has  unique document titles as docids. For MS 300K which might lack of titles, we use a document title generator \cite{title-gen} to generate high-quality titles for documents. 

    \item SEAL \cite{seal} uses n-grams as docids, and generates docids based on FM-index. It uses BART-large as the backbone.
    \item DSI-QG \cite{zhuang2022bridgingdsiqg} generates pseudo-queries conditioned on the document using docT5query~\cite{doct5query} and pairs them with docids for training. It uses unique integer strings as identifiers.
    \item NCI \cite{NCI} employs semantic structured numbers as identifiers. It trains the model using pairs of pseudo-queries and docids, and designs a prefix-aware decoder.
    \item Ultron \cite{zhou2022ultron} employs the product quantization code as docids. It starts with pre-training using document piece-docid pairs, followed by supervised fine-tuning with annotated queries and generated pseudo-queries on downstream tasks.
    \item Corpusbrain \cite{chen2022corpusbrain} employs unique document titles as docids for Wikipedia during pre-training. For MS MARCO, it might lack of titles; hence, we use the document title generator \cite{title-gen} to generate titles for documents. It undergoes pre-training using pseudo-queries constructed from documents. 
    For downstream MS 300K, we also generate document titles as docids, and then undergoes fine-tuning on downstream tasks using annotated queries.

    \item GenRet \cite{sun-2023-learning-arxiv} introduces an autoencoder to generate identifiers for documents. This autoencoder learns to compress documents into docids and to reconstruct docids back into documents. It learns jointly with the retrieval task.
    \item NOVO \cite{wang2023novo} selects important words from the document as docids. The model is trained through supervised learning with annotated information.
  
\end{enumerate*}
All GR baselines are optimized with an encoder-decoder architecture using MLE. 

\subsection{Additional Implementation Details}
\label{appendix-implementation}
For T5-base, the hidden size is 768, the feed-forward layer size is 12, the number of self-attention heads is 12, and the number of transformer layers is 12. Decoder-only structures like the GPT \cite{2022Traininggpt} series models are left for future exploration. 

BootRet and the reproduced baselines are implemented with PyTorch 1.9.0 and HuggingFace transformers 4.16.2; we re-implement DSI, and utilize open-sourced code for other baselines. 

For data augmentation during fine-tuning,  we leverage the pre-trained model, DocT5Query \cite{doct5query} to generete pseudo-queries for documents. For MS 300K, we directly use the off-the-shelf pseudo-queries \cite{doct5querygit}.
For NQ, we use the labeled queries to fine-tune DocT5Query. For each document, we generate 10 queries with the first 512 tokens of the document as input and constrain the maximum length of the generated query as 64.
During training, we pair these pseudo-queries with docids corresponding to the document, and learn these pairs with standard MLE.

\begin{table}[t]
    \centering
    \setlength{\tabcolsep}{2.8pt}
    \renewcommand{\arraystretch}{1}
    \begin{tabular}{lcccc}
        \toprule
        \textbf{Method} &  \textbf{MRR@20} &  \textbf{Hits@10}  \\
        \midrule     
        Ultron-PQ &38.41 &	62.85 \\   
        BootRet-Mt$^{MS}$ &42.79 &	66.73\\       
        monoBERT & \textbf{46.83}\rlap{$^{\ast}$}	&\textbf{71.88}\rlap{$^{\ast}$}\\
        \bottomrule
    \end{tabular}
    \caption{Comparison between GR methods and the full-ranking baseline on MS 300K. Best results are shown in \textbf{Bold}. $\ast$ indicates statistically significant improvements over BootRet  ($p \leq 0.05$).}
    \label{tab:full-ranking}
    
\end{table}

\subsection{Additional Comparisons}\label{appendix-full-ranking}
As depicted in Table \ref{tab:full-ranking}, for evaluating current GR methods against full-ranking methods, we adopt a cross-encoder baseline, namely monoBERT \cite{nogueira2019multi}. Firstly, BM25 retrieves the top 1000 candidate documents, and monoBERT subsequently ranks them. monoBERT concatenates the query and document as input, and utilizes [CLS] for relevance calculation. It is optimized with cross-entropy.

\subsection{Case Study}\label{appendix-case}
To better explain the changes in identifiers over bootstrapping iterations, we conducted a case study. Specifically, we sampled two documents. Below are their PQs at the initial stage, after training one round (BootRt-Bs), and after training multiple rounds (BootRt-Mt).

As shown in Table~\ref{tab:case-study}, we found that as identifiers evolve, the PQs for semantically similar documents gradually become more discriminative, while still maintaining appropriate similarity. This makes the semantic hierarchy of the docid prefix tree clearer.

\subsection{Inference Efficiency}\label{appendix-efficiency}
Since inference efficiency is critical for practical use,
we further evaluate memory costs and inference speed on MS 300K.
Table~\ref{tab:inference} in the Appendix highlights BootRet's significant reduction in memory and latency compared to DPR. BootRet only needs a prefix tree for inference, resulting in 93\% less memory usage than DPR's index based on dense vectors. 
Additionally, BootRet outperforms DPR, with latency dropping from 18.35ms to 8.87ms for a 300K corpus. While ANN methods speed up, dual encoder latency may increase with larger corpora. However, the inference speed of BootRet only depends on the prefix tree's structure.

\begin{table}[t]
    \centering
    \setlength{\tabcolsep}{4.8pt}
    \renewcommand{\arraystretch}{1}
    \begin{tabular}{l cc}
    
        \toprule
        \textbf{Method} &  \textbf{Memory} &  \textbf{Latency}  \\
        \midrule  
        DocT5query &\phantom{00}3.76 MB	&\phantom{0}5.61 ms\\
        DPR &940.00 MB	&18.35 ms\\
        BootRet-Bs$^\mathit{Wiki}$ &\phantom{0}65.40 MB	&\phantom{0}8.87 ms\\       
        \bottomrule
    \end{tabular}
    \caption{Results about inference efficiency on MS 300K.}
    \label{tab:inference}
    \vspace{-4mm}
\end{table}

\begin{table*}[t]
\centering
\begin{tabular}{|p{3cm}|p{3cm}|p{3cm}|p{3cm}|}
\hline
\multicolumn{1}{|c|}{Original index} & \multicolumn{1}{c|}{Initial PQ} & \multicolumn{1}{c|}{PQ obtained with BootRt-Bs} & \multicolumn{1}{c|}{PQ obtained with BootRt-Mt} \\
\hline

D2169186
(topic: Germany Gasoline Prices)&	12-45-67-11-4-56-2-21-53-67-1-8-5-42-13-53-64-78-120-63-4-113-2-4	&12-45-67-11-4-56-2-21-53-67-1-8-5-42-13-53-61-72-115-67-8-121-8-9	&12-46-70-12-4-56-2-24-53-67-1-8-5-42-13-53-61-72-115-67-8-121-8-9\\

\hline
D3126635 (topic: Heating oil average prices in Germany)	&12-45-70-11-4-56-2-21-53-22-1-8-5-42-13-53-73-78-127-56-4-113-2-4	&12-47-70-11-4-56-2-21-53-22-1-8-5-45-13-53-73-78-127-56-4-110-1-2	&12-52-79-9-4-56-2-21-53-22-1-8-5-45-13-53-73-78-127-56-4-110-1-2\\

\hline
\end{tabular}
\caption{Two sampled documents and their corresponding initial state PQ, obtained with BootRet-Bs and BootRet-Mt respectively.}
\label{tab:case-study}
\end{table*}

%% file: main.bbl
\begin{thebibliography}{54}
\expandafter\ifx\csname natexlab\endcsname\relax\def\natexlab#1{#1}\fi

\bibitem[{Anderson and Bower(2014)}]{anderson2014human}
John~R. Anderson and Gordon~H. Bower. 2014.
\newblock \emph{Human Associative Memory}.
\newblock Psychology press.

\bibitem[{Anserini()}]{Anserini}
Anserini. 2020.
\newblock Anserini.
\newblock \url{https://github.com/castorini/anserini}.

\bibitem[{Bevilacqua et~al.(2022)Bevilacqua, Ottaviano, Lewis, Yih, Riedel, and Petroni}]{seal}
Michele Bevilacqua, Giuseppe Ottaviano, Patrick Lewis, Wen-tau Yih, Sebastian Riedel, and Fabio Petroni. 2022.
\newblock Autoregressive search engines: Generating substrings as document identifiers.
\newblock In \emph{Advances in Neural Information Processing Systems}, pages 31668--31683.

\bibitem[{Chakrabarty et~al.(2018)Chakrabarty, Alhindi, and Muresan}]{chakrabarty2018robust}
Tuhin Chakrabarty, Tariq Alhindi, and Smaranda Muresan. 2018.
\newblock Robust document retrieval and individual evidence modeling for fact extraction and verification.
\newblock In \emph{Proceedings of the First Workshop on Fact Extraction and VERification}, pages 127--131.

\bibitem[{Chen et~al.(2023)Chen, Zhang, Guo, de~Rijke, Chen, Fan, and Cheng}]{chen2023-continual}
Jiangui Chen, Ruqing Zhang, Jiafeng Guo, Maarten de~Rijke, Wei Chen, Yixing Fan, and Xueqi Cheng. 2023.
\newblock Continual learning for generative retrieval over dynamic corpora.
\newblock In \emph{Proceedings of the 32nd {ACM} Conference on Information and Knowledge Management}, pages 306--315.

\bibitem[{Chen et~al.(2022)Chen, Zhang, Guo, Liu, Fan, and Cheng}]{chen2022corpusbrain}
Jiangui Chen, Ruqing Zhang, Jiafeng Guo, Yiqun Liu, Yixing Fan, and Xueqi Cheng. 2022.
\newblock Corpusbrain: Pre-train a generative retrieval model for knowledge-intensive language tasks.
\newblock In \emph{Proceedings of the 31st ACM International Conference on Information \& Knowledge Management}, pages 191--200.

\bibitem[{De~Cao et~al.(2021)De~Cao, Izacard, Riedel, and Petroni}]{genre}
Nicola De~Cao, Gautier Izacard, Sebastian Riedel, and Fabio Petroni. 2021.
\newblock Autoregressive entity retrieval.
\newblock In \emph{International Conference on Learning Representations}.

\bibitem[{Deepika and Geetha(2021)}]{deepika2021pattern}
S.S. Deepika and T.V. Geetha. 2021.
\newblock Pattern-based bootstrapping framework for biomedical relation extraction.
\newblock \emph{Engineering Applications of Artificial Intelligence}, 99:104130.

\bibitem[{Gao and Callan(2022)}]{gao2021unsupervisedcocondenser}
Luyu Gao and Jamie Callan. 2022.
\newblock Unsupervised corpus aware language model pre-training for dense passage retrieval.
\newblock In \emph{Proceedings of the 60th Annual Meeting of the Association for Computational Linguistics}, pages 2843--2853.

\bibitem[{Ge et~al.(2013)Ge, He, Ke, and Sun}]{ge2013optimized}
Tiezheng Ge, Kaiming He, Qifa Ke, and Jian Sun. 2013.
\newblock Optimized product quantization.
\newblock \emph{IEEE transactions on pattern analysis and machine intelligence}, 36(4):744--755.

\bibitem[{Jin et~al.(2023)Jin, Zeng, Wang, Chen, Wei, Li, Wang, Li, Li, Lu, Wang, Han, and Tang}]{Jin2023LanguageMA}
Bowen Jin, Hansi Zeng, Guoyin Wang, Xiusi Chen, Tianxin Wei, Ruirui Li, Zhengyang Wang, Zheng Li, Yang Li, Hanqing Lu, Suhang Wang, Jiawei Han, and Xianfeng Tang. 2023.
\newblock \href {https://api.semanticscholar.org/CorpusID:263909224} {Language models as semantic indexers}.
\newblock \emph{ArXiv}, abs/2310.07815.

\bibitem[{Karpukhin et~al.(2020{\natexlab{a}})Karpukhin, O{\u{g}}uz, Min, Lewis, and Wu}]{karpukhin2020dense}
Vladimir Karpukhin, Barlas O{\u{g}}uz, Sewon Min, Patrick Lewis, and Wu. 2020{\natexlab{a}}.
\newblock Dense passage retrieval for open-domain question answering.
\newblock In \emph{Proceedings of the 2020 Conference on Empirical Methods in Natural Language Processing}, pages 6769--6781.

\bibitem[{Karpukhin et~al.(2020{\natexlab{b}})Karpukhin, O{\u{g}}uz, Min, Lewis, Wu, Edunov, Chen, and Yih}]{pipeline3}
Vladimir Karpukhin, Barlas O{\u{g}}uz, Sewon Min, Patrick Lewis, Ledell Wu, Sergey Edunov, Danqi Chen, and Wen-tau Yih. 2020{\natexlab{b}}.
\newblock Dense passage retrieval for open-domain question answering.
\newblock In \emph{Proceedings of the 2020 Conference on Empirical Methods in Natural Language Processing}, pages 6769--6781.

\bibitem[{Khosla et~al.(2020)Khosla, Teterwak, Wang, Sarna, Tian, Isola, Maschinot, Liu, and Krishnan}]{khosla2020supervised}
Prannay Khosla, Piotr Teterwak, Chen Wang, Aaron Sarna, Yonglong Tian, Phillip Isola, Aaron Maschinot, Ce~Liu, and Dilip Krishnan. 2020.
\newblock Supervised contrastive learning.
\newblock \emph{Advances in Neural Information Processing Systems}, 33:18661--18673.

\bibitem[{Kounios et~al.(2001)Kounios, Smith, Yang, Bachman, and D'Esposito}]{kounios2001cognitive}
John Kounios, Roderick~W Smith, Wei Yang, Peter Bachman, and Mark D'Esposito. 2001.
\newblock Cognitive association formation in human memory revealed by spatiotemporal brain imaging.
\newblock \emph{Neuron}, 29(1):297--306.

\bibitem[{Kwiatkowski et~al.(2019)Kwiatkowski, Palomaki, Redfield, Collins, and Parikh}]{naturalquestion}
Tom Kwiatkowski, Jennimaria Palomaki, Olivia Redfield, Michael Collins, and Ankur Parikh. 2019.
\newblock Natural questions: A benchmark for question answering research.
\newblock \emph{Transactions of the Association for Computational Linguistics}, 7:452--466.

\bibitem[{Lee et~al.(2019)Lee, Chang, and Toutanova}]{DBLP:conf/acl/LeeCT19}
Kenton Lee, Ming{-}Wei Chang, and Kristina Toutanova. 2019.
\newblock Latent retrieval for weakly supervised open domain question answering.
\newblock In \emph{Proceedings of the 57th Conference of the Association for Computational Linguistics, {ACL} 2019, Florence, Italy, July 28- August 2, 2019, Volume 1: Long Papers}, pages 6086--6096.

\bibitem[{Lewis et~al.(2019)Lewis, Liu, Goyal, Ghazvininejad, Mohamed, Levy, Stoyanov, and Zettlemoyer}]{Lewis2019BARTDS}
Mike Lewis, Yinhan Liu, Naman Goyal, Marjan Ghazvininejad, Abdelrahman Mohamed, Omer Levy, Veselin Stoyanov, and Luke Zettlemoyer. 2019.
\newblock {BART}: Denoising sequence-to-sequence pre-training for natural language generation, translation, and comprehension.
\newblock In \emph{Proceedings of the 58th Annual Meeting of the Association for Computational Linguistics}, pages 7871--7880.

\bibitem[{Li et~al.(2023)Li, Yang, Wang, Wei, and Li}]{li2023multiview}
Yongqi Li, Nan Yang, Liang Wang, Furu Wei, and Wenjie Li. 2023.
\newblock Multiview identifiers enhanced generative retrieval.
\newblock In \emph{61st Annual Meeting of the Association for Computational Linguistics}, pages 6636--6648.

\bibitem[{Mack et~al.(2016)Mack, Love, and Preston}]{mack2016dynamic}
Michael~L Mack, Bradley~C Love, and Alison~R Preston. 2016.
\newblock Dynamic updating of hippocampal object representations reflects new conceptual knowledge.
\newblock \emph{Proceedings of the National Academy of Sciences}, 113(46):13203--13208.

\bibitem[{Mehta et~al.(2023)Mehta, Gupta, Tay, Dehghani, Tran, Rao, Najork, Strubell, and Metzler}]{mehta2022dsi++}
Sanket~Vaibhav Mehta, Jai Gupta, Yi~Tay, Mostafa Dehghani, Vinh~Q Tran, Jinfeng Rao, Marc Najork, Emma Strubell, and Donald Metzler. 2023.
\newblock {DSI++}: Updating transformer memory with new documents.
\newblock In \emph{Proceedings of the 2023 Conference on Empirical Methods in Natural Language Processing}, pages 8198--8213.

\bibitem[{Metzler et~al.(2021)Metzler, Tay, Bahri, and Najork}]{modelBased}
Donald Metzler, Yi~Tay, Dara Bahri, and Marc Najork. 2021.
\newblock Rethinking search: Making domain experts out of dilettantes.
\newblock \emph{SIGIR Forum}, 55(1):1--27.

\bibitem[{Nguyen et~al.(2016)Nguyen, Rosenberg, Song, Gao, Tiwary, Majumder, and Deng}]{msmarco}
Tri Nguyen, Mir Rosenberg, Xia Song, Jianfeng Gao, Saurabh Tiwary, Rangan Majumder, and Li~Deng. 2016.
\newblock {MS} {MARCO:} {A} human generated machine reading comprehension dataset.
\newblock In \emph{CoCo@NIPS2016}.

\bibitem[{Nie et~al.(2020)Nie, Zhang, Geng, Ramamurthy, Song, and Jiang}]{nie2020dc}
Ping Nie, Yuyu Zhang, Xiubo Geng, Arun Ramamurthy, Le~Song, and Daxin Jiang. 2020.
\newblock {DC-BERT}: Decoupling question and document for efficient contextual encoding.
\newblock In \emph{Proceedings of the 43rd International ACM SIGIR Conference on Research and Development in Information Retrieval}, page 1829–1832. ACM.

\bibitem[{Nogueira and Lin(2019{\natexlab{a}})}]{doct5querygit}
Rodrigo Nogueira and Jimmy Lin. 2019{\natexlab{a}}.
\newblock Doct5query.
\newblock \url{https://github.com/castorini/docTTTTTquery}.

\bibitem[{Nogueira and Lin(2019{\natexlab{b}})}]{doct5query}
Rodrigo Nogueira and Jimmy Lin. 2019{\natexlab{b}}.
\newblock From doc2query to doctttttquery.
\newblock An MS MARCO Passage Retrieval Task Publication.
\newblock University of Waterloo.

\bibitem[{Nogueira et~al.(2019)Nogueira, Yang, Cho, and Lin}]{nogueira2019multi}
Rodrigo Nogueira, Wei Yang, Kyunghyun Cho, and Jimmy Lin. 2019.
\newblock Multi-stage document ranking with {BERT}.
\newblock \emph{arXiv preprint arXiv:1910.14424}.

\bibitem[{Olivares et~al.(2023)Olivares, Quijano, and Liberatore}]{olivares2023enhancing}
Daniel~Guzman Olivares, Lara Quijano, and Federico Liberatore. 2023.
\newblock Enhancing information retrieval in fact extraction and verification.
\newblock In \emph{Proceedings of the Sixth Fact Extraction and VERification Workshop}, pages 38--48.

\bibitem[{Ouyang et~al.(2022)Ouyang, Wu, Jiang, Almeida, Wainwright, Mishkin, Zhang, Agarwal, Slama, Ray et~al.}]{2022Traininggpt}
Long Ouyang, Jeffrey Wu, Xu~Jiang, Diogo Almeida, Carroll Wainwright, Pamela Mishkin, Chong Zhang, Sandhini Agarwal, Katarina Slama, Alex Ray, et~al. 2022.
\newblock Training language models to follow instructions with human feedback.
\newblock In \emph{Advances in Neural Information Processing Systems}, volume~35, pages 27730--27744.

\bibitem[{Pradeep et~al.(2023)Pradeep, Hui, Gupta, Lelkes, Zhuang, Lin, Metzler, and Tran}]{pradeep2023does-scale-genir}
Ronak Pradeep, Kai Hui, Jai Gupta, Adam~D Lelkes, Honglei Zhuang, Jimmy Lin, Donald Metzler, and Vinh~Q Tran. 2023.
\newblock How does generative retrieval scale to millions of passages?
\newblock In \emph{Gen-IR@SIGIR 2023: The First Workshop on Generative Information Retrieval}.

\bibitem[{Raffel et~al.(2020)Raffel, Shazeer, Roberts, Lee, Narang, Matena, Zhou, Li, and Liu}]{raffel2020exploringt5}
Colin Raffel, Noam Shazeer, Adam Roberts, Katherine Lee, Sharan Narang, Michael Matena, Yanqi Zhou, Wei Li, and Peter~J. Liu. 2020.
\newblock Exploring the limits of transfer learning with a unified text-to-text transformer.
\newblock \emph{The Journal of Machine Learning Research}, 21(1):5485--5551.

\bibitem[{Raffel et~al.(2021)Raffel, Shazeer, Roberts, Lee, Narang, Matena, Zhou, Li, and Liu}]{t5base}
Colin Raffel, Noam Shazeer, Adam Roberts, Katherine Lee, Sharan Narang, Michael Matena, Yanqi Zhou, Wei Li, and Peter~J. Liu. 2021.
\newblock T5 base.
\newblock \url{https://huggingface.co/t5-base}.

\bibitem[{Ren et~al.(2023)Ren, Zhao, Liu, Wu, Wen, and Wang}]{ren2023tome}
Ruiyang Ren, Wayne~Xin Zhao, Jing Liu, Hua Wu, Ji-Rong Wen, and Haifeng Wang. 2023.
\newblock {TOME}: A two-stage approach for model-based retrieval.
\newblock In \emph{Proceedings of the 61st Annual Meeting of the Association for Computational Linguistics}, pages 6102--6114.

\bibitem[{Robertson et~al.(1995)Robertson, Walker, Jones, Hancock-Beaulieu, and Gatford}]{bm25}
Stephen~E. Robertson, Steve Walker, Susan Jones, Micheline~M. Hancock-Beaulieu, and Mike Gatford. 1995.
\newblock Okapi at {TREC}-3.
\newblock In \emph{TREC}, pages 109--126.

\bibitem[{Song and Roth(2014)}]{song2014dataless}
Yangqiu Song and Dan Roth. 2014.
\newblock On dataless hierarchical text classification.
\newblock In \emph{Proceedings of the AAAI Conference on Artificial Intelligence}, volume~28, pages 1579--1585.

\bibitem[{Sun et~al.(2023)Sun, Yan, Chen, Wang, Zhu, Ren, Chen, Yin, de~Rijke, and Ren}]{sun-2023-learning-arxiv}
Weiwei Sun, Lingyong Yan, Zheng Chen, Shuaiqiang Wang, Haichao Zhu, Pengjie Ren, Zhumin Chen, Dawei Yin, Maarten de~Rijke, and Zhaochun Ren. 2023.
\newblock Learning to tokenize for generative retrieval.
\newblock In \emph{Advances in Neural Information Processing Systems}, volume~36.

\bibitem[{Tang et~al.(2023)Tang, Zhang, Guo, and de~Rijke}]{tang-2023-recent}
Yubao Tang, Ruqing Zhang, Jiafeng Guo, and Maarten de~Rijke. 2023.
\newblock Recent advances in generative information retrieval.
\newblock In \emph{SIGIR-AP 2023: 1st International ACM SIGIR Conference on Information Retrieval in the Asia Pacific}, pages 294--297. ACM.

\bibitem[{Tay et~al.(2022)Tay, Tran, Dehghani, Ni, Bahri, Mehta, Qin, Hui, Zhao, Gupta, Schuster, Cohen, and Metzler}]{DSI}
Yi~Tay, Vinh~Q. Tran, Mostafa Dehghani, Jianmo Ni, Dara Bahri, Harsh Mehta, Zhen Qin, Kai Hui, Zhe Zhao, Jai Gupta, Tal Schuster, William~W. Cohen, and Donald Metzler. 2022.
\newblock Transformer memory as a differentiable search index.
\newblock In \emph{Advances in Neural Information Processing Systems}, volume~35, pages 21831--21843.

\bibitem[{Touvron et~al.(2023)Touvron, Lavril, Izacard, Martinet, Lachaux, Lacroix, Rozi{\`e}re, Goyal, Hambro, Azhar et~al.}]{LLaMA13b}
Hugo Touvron, Thibaut Lavril, Gautier Izacard, Xavier Martinet, Marie-Anne Lachaux, Timoth{\'e}e Lacroix, Baptiste Rozi{\`e}re, Naman Goyal, Eric Hambro, Faisal Azhar, et~al. 2023.
\newblock {LLaMA}: Open and efficient foundation language models.
\newblock \url{https://huggingface.co/huggyllama/llama-13b}.
\newblock Meta AI.

\bibitem[{Wang et~al.(2022)Wang, Hou, Wang, Miao, Wu, Sun, Chen, Xia, Chi, Zhao, Liu, Xie, Sun, Deng, Zhang, and Yang}]{NCI}
Yujing Wang, Yingyan Hou, Haonan Wang, Ziming Miao, Shibin Wu, Hao Sun, Qi~Chen, Yuqing Xia, Chengmin Chi, Guoshuai Zhao, Zheng Liu, Xing Xie, Hao Sun, Weiwei Deng, Qi~Zhang, and Mao Yang. 2022.
\newblock A neural corpus indexer for document retrieval.
\newblock In \emph{Advances in Neural Information Processing Systems}, volume~35, pages 25600--25614.

\bibitem[{Wang et~al.(2023)Wang, Zhou, Tu, and Dou}]{wang2023novo}
Zihan Wang, Yujia Zhou, Yiteng Tu, and Zhicheng Dou. 2023.
\newblock {NOVO}: Learnable and interpretable document identifiers for model-based {IR}.
\newblock In \emph{Proceedings of the 32nd {ACM} Conference on Information and Knowledge Management}.

\bibitem[{Wikipedia(2022)}]{wikidump}
Wikipedia. 2022.
\newblock Data dumps.
\newblock \url{https://dumps.wikimedia.org/enwiki/latest/enwiki-latest-pages-articles.xml.bz2}.

\bibitem[{Wu et~al.(2009)Wu, Lee, Ye, and Chieu}]{wu2009domain}
Dan Wu, Wee~Sun Lee, Nan Ye, and Hai~Leong Chieu. 2009.
\newblock Domain adaptive bootstrapping for named entity recognition.
\newblock In \emph{Proceedings of the 2009 Conference on Empirical Methods in Natural Language Processing, Volume 3}, pages 1523--1532. Association for Computing Machinery.

\bibitem[{Xiong et~al.(2017)Xiong, Zhong, and Socher}]{xiongdcn+mle}
Caiming Xiong, Victor Zhong, and Richard Socher. 2017.
\newblock Dcn+: Mixed objective and deep residual coattention for question answering.
\newblock In \emph{International Conference on Learning Representations}.

\bibitem[{Xiong et~al.(2020)Xiong, Xiong, Li, Tang, Liu, Bennett, Ahmed, and Overwijk}]{xiong2020approximate}
Lee Xiong, Chenyan Xiong, Ye~Li, Kwok-Fung Tang, Jialin Liu, Paul~N Bennett, Junaid Ahmed, and Arnold Overwijk. 2020.
\newblock Approximate nearest neighbor negative contrastive learning for dense text retrieval.
\newblock In \emph{International Conference on Learning Representations}.

\bibitem[{Zearing(2023)}]{title-gen}
Caleb Zearing. 2023.
\newblock Article title generator.
\newblock \url{https://huggingface.co/czearing/article-title-generator}.

\bibitem[{Zeng et~al.(2023)Zeng, Luo, Jin, Sarwar, Wei, and Zamani}]{zeng2023scalable}
Hansi Zeng, Chen Luo, Bowen Jin, Sheikh~Muhammad Sarwar, Tianxin Wei, and Hamed Zamani. 2023.
\newblock Scalable and effective generative information retrieval.
\newblock \emph{arXiv preprint arXiv:2311.09134}.

\bibitem[{Zhan et~al.(2020{\natexlab{a}})Zhan, Mao, Liu, Guo, Zhang, and Ma}]{pipeline4}
Jingtao Zhan, Jiaxin Mao, Yiqun Liu, Jiafeng Guo, Min Zhang, and Shaoping Ma. 2020{\natexlab{a}}.
\newblock Optimizing dense retrieval model training with hard negatives.
\newblock In \emph{Proceedings of the 44th International ACM SIGIR Conference on Research and Development in Information Retrieval}, pages 1503--1512.

\bibitem[{Zhan et~al.(2021)Zhan, Mao, Liu, Guo, Zhang, and Ma}]{zhan2021jointly}
Jingtao Zhan, Jiaxin Mao, Yiqun Liu, Jiafeng Guo, Min Zhang, and Shaoping Ma. 2021.
\newblock Jointly optimizing query encoder and product quantization to improve retrieval performance.
\newblock In \emph{Proceedings of the 30th ACM International Conference on Information \& Knowledge Management}, pages 2487--2496.

\bibitem[{Zhan et~al.(2020{\natexlab{b}})Zhan, Mao, Liu, Zhang, and Ma}]{zhan2020repbert}
Jingtao Zhan, Jiaxin Mao, Yiqun Liu, Min Zhang, and Shaoping Ma. 2020{\natexlab{b}}.
\newblock {RepBERT}: Contextualized text embeddings for first-stage retrieval.
\newblock \emph{arXiv preprint arXiv:2006.15498}.

\bibitem[{Zhang et~al.(2023)Zhang, Liu, Zhou, Dou, and Cao}]{zhang2023term-sets-arxiv}
Peitian Zhang, Zheng Liu, Yujia Zhou, Zhicheng Dou, and Zhao Cao. 2023.
\newblock Term-sets can be strong document identifiers for auto-regressive search engines.
\newblock \emph{arXiv preprint arXiv:2305.13859}.

\bibitem[{Zhou et~al.(2023)Zhou, Dou, and Wen}]{zhou-2023-enhancing}
Yujia Zhou, Zhicheng Dou, and Ji-Rong Wen. 2023.
\newblock Enhancing generative retrieval with reinforcement learning from relevance feedback.
\newblock In \emph{EMNLP 2023: Proceedings of the 2023 Conference on Empirical Methods in Natural Language Processing}.

\bibitem[{Zhou et~al.(2022)Zhou, Yao, Dou, Wu, Zhang, and Wen}]{zhou2022ultron}
Yujia Zhou, Jing Yao, Zhicheng Dou, Ledell Wu, Peitian Zhang, and Ji-Rong Wen. 2022.
\newblock Ultron: An ultimate retriever on corpus with a model-based indexer.
\newblock \emph{arXiv preprint arXiv:2208.09257}.

\bibitem[{Zhuang et~al.(2023)Zhuang, Ren, Shou, Pei, Gong, Zuccon, and Jiang}]{zhuang2022bridgingdsiqg}
Shengyao Zhuang, Houxing Ren, Linjun Shou, Jian Pei, Ming Gong, Guido Zuccon, and Daxin Jiang. 2023.
\newblock Bridging the gap between indexing and retrieval for differentiable search index with query generation.
\newblock In \emph{Gen-IR@SIGIR 2023: The First Workshop on Generative Information Retrieval}.

\end{thebibliography}
